\documentclass[proof]{WileyASNA-v1}

\articletype{Article Type}%

\received{xx April 2022}
\revised{xx June 2022}
\accepted{xx June 2022}

\begin{document}

\title{Probing the Przybylski star for deuterium}

\author[1,2,3]{Sergei M. Andrievsky*}

\author[1,2]{Valery V. Kovtyukh}

\authormark{Sergei M. Andrievsky \& Valery V. Kovtyukh} 

\address[1]{\orgdiv{Astronomical Observatory}, \orgname{Odessa National University of the Ministry of 
Education and Science of Ukraine}, \orgaddress{\state{Shevchenko Park, 65014, Odessa}, \country{Ukraine}}}

\address[2]{\orgdiv{Institut f\"{u}r Astronomie und Astrophysik}, \orgname{Kepler Center for 
Astro and Particle Physics, Universit\"{a}t T\"{u}bingen}, \orgaddress{\state{Sand 1, 72076 T\"{u}bingen}, \country{Germany}}}

\address[3]{\orgdiv{GEPI}, \orgname{Observatoire de Paris, Universit\'e PSL, CNRS}, 
\orgaddress{\state{ 5 Place Jules Janssen, F-92190 Meudon}, \country{France}}}

\corres{*S.~M.~Andrievsky, Astronomical Observatory, Odessa National University of the Ministry of Education and Science
of Ukraine, Shevchenko Park, 65014, Odessa, Ukraine \email{andrievskii@ukr.net}}

\abstract{The Przybylski star spectrum has been studied in order to search for deuterium lines.
Since this star is extremely enriched in the $s$-process elements, which are the product 
of interaction between free neutrons and seed nuclei, we might as well expect to detect
deuterium in this star. However, no visible spectroscopic manifestation of deuterium has been
detected. Perhaps, the reason of this result is the convective "destruction" of this isotope.}



\maketitle

\section{Introduction}

The Przybylski star (HD~101065) is a unique object of our Galaxy. A recent very detailed description 
of all the peculiarities of this star was given in the paper of \cite{Andrievsky2022}. In that paper, 
a previous scenario published by \cite{GopUlAndr2008} was further developed that enabled one to 
explain some peculiarities of this enigmatic star. It was proposed that this star is a component
of a binary system with a companion neutron star which is a $\gamma$-ray pulsar for it. This hypothesis
allows one to naturally overcome that obstacle that we do not register orbital motion of the
Przybylski star in the spectroscopic observations. The $\gamma$-ray pulsar, being the source of 
the power $\gamma$-radiation emitted approximately in the orthogonal direction to the magnetic axis,
irradiates the atmosphere of this star. High-energy $\gamma$-quanta interacting with the nuclei
of quite abundant species of the gas of the Przybylski's star atmosphere can produce so-called
photoneutrons, which are quickly thermalized. As resonant neutrons, they interact with the
seed nuclei (e.g., such as iron, nickel and others) thus producing in this way $s$-process elements. 
All the details of this scenario can be found in the aforementioned paper of \cite{Andrievsky2022}.
The author also raises another question, which does not seem to have been discussed in the 
literature before. It is as follows: if we are detecting the enormous amount of the $s$-process 
elements in the atmosphere of this star, which are undoubtedly produced only in the interactions 
with free neutrons, then should we also detect the deuterium in the Przybylski star? This question 
is $independent$ of the validity of the proposed hypothesis. Deuterium must be produced due to 
the neutrons capture by the hydrogen nuclei. Really, as it is shown by \cite{Mughabghab2003} and 
\cite{PritMugh2012} the cross-section of the interaction between thermalized neutrons and protons 
is high, about 0.3 barn (for the next quite abundant nuclei, like C, N, O the corresponding value 
is much smaller). This means that the yield of deuterium in the Przybylski's star envelope does 
not look completely impossible. In the following sections we will discuss this problem in detail.  

\section{Galactic evolution of deuterium}

It is know that deuterium nuclei were mainly produced in the Big Bang nucleosynthesis. 
\cite{Epsteinetal1976} considered all possible processes of free neutron production and 
possible source of the deuterons formation outside the Big Bang nucleosynthesis. Among 
the possible sources, the authors considered spallation reactions (for example, interactions 
between energetic protons and $\alpha$-particles), pre-galactic cosmic rays (protons 
and  $\alpha$-particles) interactions with hydrogen and helium, shock waves propagating 
through low-density gas, hot explosions of different kinds, disrupted neutron stars (e.g., 
as a result of their mergers with black holes). Having considered all these mechanisms 
the authors concluded that the most reasonable place of origin for deuterium is Big Bang 
nucleosynthesis. Of course, this is true if we are talking about the global production of this 
element in the Universe. However, local overproduction of deuterium is not excluded. For 
example, \cite{MullLyn1998}, \cite{ProdField2003} considered non-primordial deuterium 
production by accelerated particles born in the flares of dwarf stars. 

\section{Spectroscopic manifestation of the deuterium presence in the stellar atmosphere}

When deuterons are born from the interaction between neutrons and protons, 
they form deuterium component of the gas in the stellar atmosphere. Therefore, 
we must be able to spectroscopically detect the deuterium lines. Such a task
was undertaken in the present paper.

To calculate the wavelengths of the deuterium lines (blue-shifted compared to the 
hydrogen lines), we will use the following formula:

$\Delta \lambda \approx \lambda_{\rm H} \frac{m_{\rm e}}{2 m_{\rm p}}.$

Here $\Delta \lambda$ is a difference between the wavelenghts of the deuterium 
and hydrogen lines, $m_{\rm e}$ and $m_{\rm p}$ are the masses of the electron 
and proton, respectively. Table 1 lists the $\Delta \lambda$ values for the 
several Balmer and Paschen lines.

\begin{table}
\caption{Wavelength shifts between hydrogen and deuterium lines}
\label{Line}
\begin{tabular}{lc}
\hline
Line, \AA  & $\Delta \lambda$, \AA \\  
\hline
H$_\alpha$      &    --1.78   \\
H$_\beta$       &    --1.32   \\
H$_\gamma$      &    --1.18   \\
\hline
P$_\alpha$     &    --5.10   \\ 
P$_\beta$      &    --3.48   \\ 
\hline
\end{tabular}
\end{table}

It should be noted that several spectra of the Przybylski star were obtained with 
$IUE$ near the Ly$\alpha$ line, and they are available from archive. Since this emission 
line is rather broad, and the estimated shift between the hydrogen and deuterium 
lines is only about 0.3 \AA, there is no way to separate these lines and to identify the
deuterium line. Therefore, we have concentrated only on the lines of the Balmer and Paschen
series.

\begin{figure} 
\resizebox{\hsize}{!}{\includegraphics{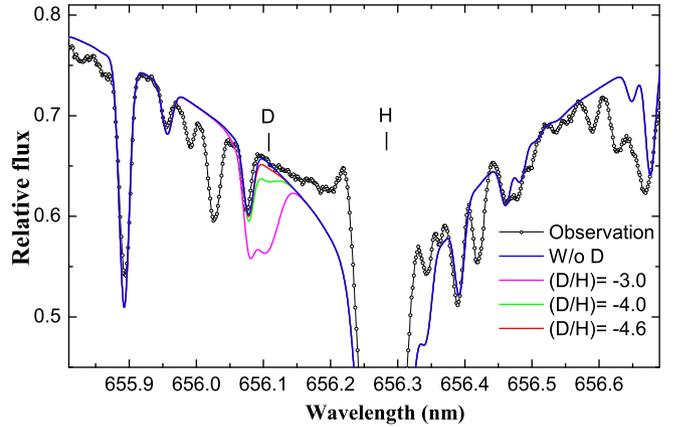}}    
\caption{Fragment of the observed spectrum of the Przybylski star, covering
the H$\alpha$ line (solid line with open circles) and four variants of the 
synthetic spectra: without deuterium (blue), relative deuterium abundance 
[D/H] = --3 (magenta), --4.0 (green), --4.6 (red)}
\label{H_Alpha_D}
\end{figure}

Fig. \ref{H_Alpha_D} shows the synthetic and observed ($HARPS$, R = 110\,000, VLT ESO archive) 
spectra in the vicinity of the H$\alpha$ line. The synthetic spectrum was calculated using the 
stellar atmosphere model generated using the LLMODELS code, the atmosphere parameters and the 
chemical composition of the Przybylski's star atmosphere obtained in \cite{Shulyaketal2010}. 
These authors emphasize that rare-earth elements play a key role in the radiative energy 
balance in the atmosphere of this star, and this situation significantly differs from the case 
of ordinary stars. To calculate the hydrohen line profiles, the synthetic spectrum code by 
\cite{Tsymbaletal2019} was used. The NLTE effects were taken into account (MULTI code 2.3, 
\citealt{Carlsson1986}; hydrogen atomic model of \citealt{Mashonkina2008}). Note that 
the calculations cannot reproduce the observed profile of H$\alpha$ line in the line core 
and near wings. This effect can also be seen in the Fig. 1 from \cite{Shulyaketal2010}.
The line core is too narrow, and in addition, the presence of the blue-shifted emission-like 
feature can be suspected.

\begin{figure} 
\resizebox{\hsize}{!}{\includegraphics{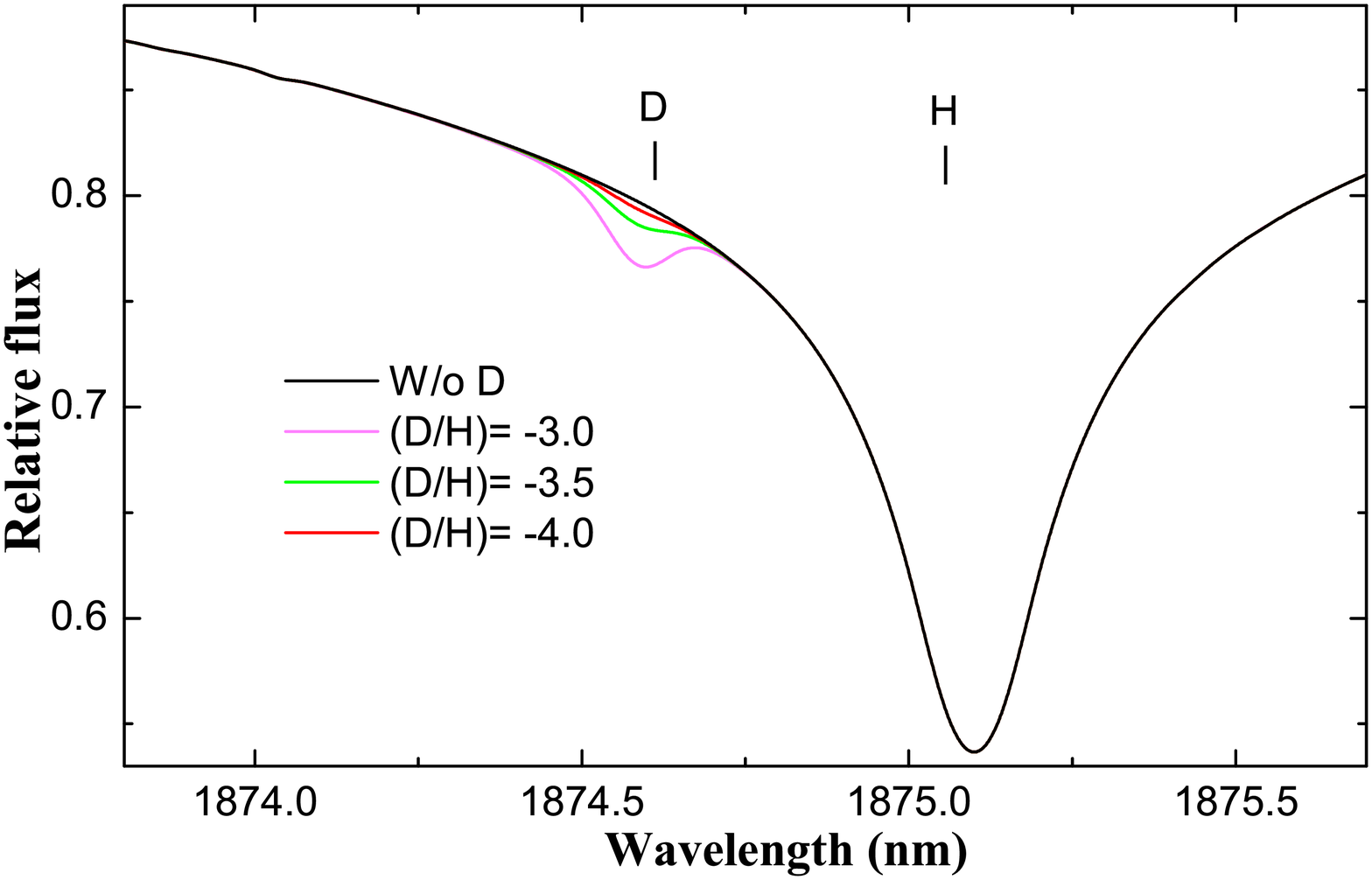}}    
\caption{Same as Fig. \ref{H_Alpha_D} but for D(P)$\alpha$}
\label{Pashen_Alpha}
\end{figure}

The deuterium D(H)$\alpha$ line profile was calculated with three adopted relative abundances:
[D/H] = --3, --4 and --4.6. Fig. \ref{H_Alpha_D} shows the calculated profiles of the deuterium Balmer 
$\alpha$ line for these three adopted deuterium abundances. Note that already for D(H)$\beta$, the 
calculated line profiles are too weak, and they are not shown here. Fig. \ref{Pashen_Alpha} shows the 
same for Paschen $\alpha$ line (only calculated profiles are shown, since there are no observed spectra 
of this star in the corresponding region). The profile of the deuterium Paschen$\beta$ line is also 
very weak.

From Fig. \ref{H_Alpha_D} one can conclude that the relative deuterium abundance  
in the Przybylski's star atmosphere must be lower than --4 dex (recall that the cosmic
abundance of this isotope (D/H) is about $10^{-5}$; recently \citealt{Weinberg2017} obtained that the 
relative abundance of the deuterium in the interstellar medium (D/H) is $1 - 1.5 \times 10^{-5}$).
 
\section{Discussion and concluding remarks}

Why do we see no signs of increased deuterium  abundance in the Przybylski's star atmosphere? 
The simplest solution is to consider convection in the upper layers of the late main-sequence
stars. For example, the Sun's convective zone has a length of about 30\% of its radius.
Here, the inner boundary of convective zone borders with the radiative zone. The temperature 
at the bottom of convective zone exceeds $10^{6}$ K. At this temperature, the deuterons can 
be "destroyed" in fusion reactions in the following way:

$ {}^{1}_{1}H + {}^{2}_{1}H = {}^{3}_{2}He + \gamma$

Overshooting must exist in the stars, otherwise it is impossible to reproduce
some observational facts related to surface abundance of some chemical elements 
in the main-sequence stars. For example, the standard solar model cannot 
explain the observed abundance of lithium, because the surface material is 
unable to reach layers having a temperature sufficient for effective reactions 
between lithium nuclei and protons. \cite{SchlatWeiss1999} bring into 
consideration additional mixing at the bottom of the convective zone in order 
to solve the problem. A very detailed classification of the processes leading 
to the phenomenon under discussion can be found in \cite{Andrassy2015}.

We do not know the internal structure of the Przybylski star, especially the characteristics 
of its convective zone: its extent, temperature at the bottom etc. These characteristics may 
be quite different from those of an ordinary late main-sequence F-star because of the extremely 
high line blanketing of this star's atmosphere. Therefore, for our purpose, to have some qualitative 
estimates, we use the necessary data for the Sun (the difference in effective temperature with the Przybylski 
star is about 850 K; the T$_{\rm eff}$ value for the Przybylski star is taken from \citealt{Mkrtichianetal2008}).

\cite{XioDen2002}, using non-local convection theory for chemically inhomogeneous stars, studied the solar 
overshooting zone based on observed lithium abundance. The authors showed that convection can penetrate 
layers with the temperature around $2.2-2.3 \times 10^{6}$ K, while the size of this zone is about 30\% 
of the radius (see their Table 1).
 
If standard convection and overshooting can bring surface material to the layers having the temperature 
of about $2.2-2.3$ MK, then the reaction $^{2}_{1}H({}^{1}_{1}H,\gamma)^{3}_{2}He$ at this temperature 
has a rate about $5 \times 10^{-8}$ cm$^{3}$~mol$^{-1}$~s$^{-1}$ (see analytical relation from 
\citealt{Anguloetal1999}). For the center of the Sun, the corresponding value is about 
$10^{-2}$ cm$^{3}$~mol$^{-1}$~s$^{-1}$, and characteristic reaction time is about of a few seconds. 
Therefore, to have some qualitative result we can roughly estimate the reaction characteristic time 
at the 2 MK as $10^{5}-10^{6}$~s, which is on the order of days. The time is taken for the stellar 
material to return from beneath the convective zone to the photosphere (a distance of about 
$3 \times 10^{10}$ cm; according to \citealt{Kurtz1980} radius of the Przybylski star is 
1.3 R$_{\odot}$) should be not less than this characteristic time. In such a case, the returned material 
will be depleted of deuterium, while the elements of $s$-process will remain unchanged. As it was 
mentioned above, the $s$-process element production was considered in \cite{Andrievsky2022} 
as a result of reactions of the seed nuclei with free thermalized photoneutrons, which are formed 
by the high-energy $\gamma$-quanta penetrating into the deep layers, where their free path becomes 
quite small (for instance, at the bottom of the convective zone the density is about 
$10^{-1}$ g~cm$^{-1}$; free path is only about $10^{5}$ cm, thus the neutron concentration may be high). 

An additional interesting question arises in connection with the above discussion. 
If there are no visible deuterium lines in the program star spectra, which means that there is no 
noticeable deuterium content in the Przybylski star's atmosphere, then how does this compare to 
the presence of other light nuclei, say, lithium nuclei? According to the works of \cite{WegPet1974} 
and \cite{Shavrinaetal2003}, curves-of-growth and synthetic spectrum analyses show that this star
possesses a normal atmosphere lithium abundance, about 3 dex in the scale where log(H)= 12.00
(see Fig. 1 from \citealt{Ramirezetal2012}). 

Let us estimate the characteristic reaction time at the base of the convective zone, where
lithium nuclei (interacting with protons) can be converted into $\alpha$-particles.
Similarly to estimate made above, we obtain the rate of the reaction $^{7}_{3}Li({}^{1}_{1}H,^{4}_{2}He)^{4}_{2}He$ 
at the center of the Sun $10^{-5}$ cm$^{3}$~mol$^{-1}$~s$^{-1}$ (\citealt{Anguloetal1999}), which 
corresponds to about twenty minutes. This value is about thirteen order of magnitudes 
greater than for a temperature of about 2.2 MK. Thus, lithium nuclei are able to remain 
in the convective zone and atmosphere for a long time.

Summarizing,

1. Deuterium atoms can be formed in the atmosphere of Przybylski star (and they are 
most likely formed there) in reactions between free neutrons and hydrogen nuclei.

2. Due to the convective circulation in the outer convective zone, the material from
the upper layers reaches the bottom of the convective zone, where the temperature
exceeds $10^{6}$~K, and there deuterons react with protons. The material returned to the 
photosphere is depleted in deuterium.

3. We have found no spectroscopic confirmation of the presence of deuterium H$\alpha$
line in the spectrum of Przybylski star. The upper limit of the relative deuterium 
abundance [D/H] is less than --4 dex.

\subsection*{acknowledgements}

SMA and VVK are grateful to the \emph{Vector-Stiftung at Stuttgart, Germany}, for support within the 
program "2022--Immediate help for Ukrainian refugee scientists" under grants P2022-0063 and P2022-0064. 
Especial thanks to Prof. K.~Werner and Dr. V.~Suleimanov for their help with organizing our stay
in the Institute of Astronomy and Astrophysics of the T\"ubingen University. This work is based on 
data collected at ESO VLT (prog. 68.D-0254, 194.C-0833, 076.D-0169, 70.D-0470, 165.N-0276) 
and HARPS (prog. 072.D-0286). Special thanks to Dr. S.~A.~ Korotin for the profile calculation.
We are extremely grateful for the referee, whose comments strengthened the conclusions of this work.

\subsection*{Conflict of interest}

The authors declare no potential conflict of interests.

\end{document}